\documentclass[aoas,preprint]{imsart}

\RequirePackage[OT1]{fontenc}
\RequirePackage{amsthm,amsmath}
\RequirePackage{natbib}
\RequirePackage[colorlinks,citecolor=blue,urlcolor=blue]{hyperref}

\usepackage{graphicx}
\usepackage{mathtools}
\usepackage{lineno}

\arxiv{arXiv:1403.5196}

\startlocaldefs
\numberwithin{equation}{section}
\theoremstyle{plain}

\endlocaldefs

\begin{document}

\begin{frontmatter}
\title{Calibration of a Natural History Model of Bowel Cancer Progression using Likelihood Emulation}
\runtitle{Calibration of a Natural History Model}

\begin{aug}
\author{\fnms{Jeremy E. } \snm{Oakley}\thanksref{m1}\ead[label=e1]{j.oakley@sheffield.ac.uk}},
\and
\author{\fnms{Benjamin D.} \snm{Youngman}\thanksref{m2}
\ead[label=e3]{b.youngman@exeter.ac.uk}
}

\runauthor{J. E. Oakley and B. D. Youngman}

\affiliation{University of Sheffield\thanksmark{m1} and University of Exeter\thanksmark{m2}}

\address{School of Mathematics and Statistics\\
The Hicks Building\\
Hounsfield Road\\
Sheffield\\
S3 7RH\\
UK\\
\printead{e1}\\
}

\address{Exeter Climate Systems\\
College of Engineering, Mathematics and Physical Sciences\\
University of Exeter\\
North Park Road\\
Exeter\\
EX4 4QF\\
UK\\
\printead{e3}\\
}
\end{aug}

\begin{abstract}
We calibrate a Natural History Model, which is a class of computer simulator used in the health industry, and here has been used to characterise bowel cancer incidence for the UK. The simulator tracks the development of bowel cancer in a sample of people, and its output mostly stratifies bowel cancer occurrence by patient age and bowel cancer type. Its output relies on 25 unknown inputs, which we are required to calibrate. In order to do this we must address that not only is the output count data, but it is also stochastic, due to the simulation procedure.

We cannot feasibly achieve calibration of the simulator using Monte Carlo methods alone, as it is of `moderate' computational expense. To achieve a reliable calibration, we must also specify its discrepancy: how, when calibrated, it differs from reality. We propose a method for calibration that combines a statistical emulator for the likelihood function with importance sampling. The emulator provides an interim sample of inputs at which the simulator is run, from which the likelihood is calculated. Importance sampling is then used to re-weight the inputs and provide a final sample of calibrated inputs. Re-calculating the importance weights incurs little computational cost, and so we can easily investigate how different discrepancy specifications affect calibration. 
\end{abstract}


\begin{keyword}
\kwd{Bayesian inference}
\kwd{calibration}
\kwd{computer simulator}
\kwd{statistical emulation}
\kwd{importance sampling}
\end{keyword}

\end{frontmatter}

\section{Introduction} \label{intro}

We aim to calibrate a Natural History Model (NHM) that \cite{pt-thesis} developed to characterise UK bowel cancer incidence. Using several different types of observed `target data', we calibrate the NHM by finding input values that make the NHM's outputs match the target data as closely as possible. There are 25 calibration inputs, $X$, which are unknown; some, however, are in principal physically observable. By calibrating the NHM, we will derive the joint distribution of $X$ given the target data. An outline of the NHM's workings is given in section \ref{nhm-sim}.

The motivation for calibrating such a model is to support decision making. In the UK, the National Institute for Health and Care Excellence (NICE) regularly makes such healthcare resource allocation decisions on the basis of cost-effectiveness, with the decisions typically informed by simulator predictions (for example scenarios see \cite{tapp3}). Furthermore, NICE expects analysts to account for simulator input uncertainty, preferably by assigning probability distributions to the inputs and deriving the simulator output distributions \cite[Section 5.8.7]{nice}. The calibrated input distributions can be used for this purpose.

Our approach to calibration is inspired by the framework for Bayesian calibration for computer models (which we refer to as `simulators') proposed by \cite{ken-ohag} and developed in \cite{higdon2004}, \cite{bayarii1}, \cite{bayarii2} and \cite{higdon2008}, and by Bayes linear history matching developed in \cite{craig2001}, \cite{goldroug2006} and \cite{vern-galaxy}. Our calibration problem involves methodology to address three issues: computationally expensive simulators, `discrepancy', which is the error in a simulator prediction due to the simulator being an imperfect model of reality, and stochastic simulators, which are simulators that can return different output values when run repeatedly at the same input values.

Any calibration method will involve running the simulator at different input values, and so methods that require large numbers of simulator runs become impractical if a single simulator run at one input value takes a long time. A well-established technique for handling expensive simulators, proposed in \cite{sacks}, is to construct a cheap surrogate model or `emulator' of the simulator using Gaussian process regression, based on a relatively small number of simulator runs. Variations of this method are used in the above references.  In this paper, the simulator is of `moderate' computational cost, with a single run at one input value taking between one and two minutes. We argue that this changes the nature of the surrogate modelling problem. In our proposed approach, rather than attempting to construct a very precise emulator of the simulator, we use a cruder emulator to guide us to the appropriate regions of the input space, and then do direct simulator evaluations in those regions. In particular, we propose the use of importance sampling, where the emulator is used to construct the importance density.        

When calibrating a simulator, it is important to account for simulator discrepancy for two reasons. Firstly, if the inputs are physically meaningful quantities that could, in principle, be observed directly, calibrating a simulator without accounting for discrepancy may result in biased estimates with severe over-confidence, as demonstrated in \cite{bryn}. If the simulator inputs are `tuning' parameters that are not physically observable, discrepancy plays an important role when calibrating to multiple outputs, or when we wish to predict unobserved output quantities using a calibrated simulator. Suppose that we have a physical observation for an output quantity $Z_1$, and wish to predict an unobserved output quantity $Z_2$. A simulator input value may give a poor fit to output $Z_1$, but a good prediction of $Z_2$. If we do not believe the simulator models $Z_1$ perfectly, we would not necessarily want to rule out such an input value and corresponding prediction of $Z_2$. Accounting for the simulator error in modelling $Z_1$ would prevent this.

Accounting for simulator discrepancy is clearly important if the simulator is being used to support decision making. Without discrepancy, we may have spuriously precise input distributions, resulting in spuriously precise output predictions. Incorporating discrepancy allows decision makers to test the robustness of their decisions both directly to errors in the model outputs, and to the broader input distributions that result from the calibration. 

As argued in \cite{bryn}, it is important to specify meaningful proper prior distributions for simulator discrepancy, but to do this may be difficult. In \cite{vern2010}, within a Bayes linear framework, the simulator expert only provided an \emph{interval} for the variance of a discrepancy parameter. \cite{strong} suggest `opening the black box' and incorporating discrepancy terms within the simulator, so that the expert considers sources of simulator discrepancy explicitly, rather than attempting to make judgements about the overall discrepancy. We argue that it is desirable to be able to investigate, without too much difficulty, a range of different discrepancy distributions, within any calibration methodology. Within our proposed importance sampling framework, we suggest an initial, conservative specification of simulator discrepancy, which can then be varied with little extra computational effort via re-calculation of importance weights corresponding to different discrepancy distributions. 

The final issue we consider is that of a stochastic simulator, which raises the question of what it is we should be trying to emulate, assuming that an emulator is necessary. The simulator in our case study produces random count data. In a similar scenario, \cite{hendetal} constructed emulators for probabilities from which the count data were assumed to have been generated. Here, we propose constructing an emulator for the likelihood function given the observed data. Our simulator produces 30 count data outputs (with various dependencies between the outputs), and so emulating the likelihood reduces the computational effort to emulating a univariate output, and enables us to implement an importance sampling approach for the calibration.

This paper has the following structure. The next section outlines the calibration method. Section \ref{nhm} presents the results of calibration of a Natural History Model and section \ref{discuss} offers conclusions and discussion of the calibration method.

\section{Outline method} \label{outline}

\subsection{The calibration problem}
We have target data $Z$, observed in the real world, with which we can calibrate the simulator. The data are made up of observations of various binomial and multinomial random variables, but to simplify the discussion, we suppose that $Z$ is a single binomial random variable, with $Z|\theta^* \sim Bin(N,\theta^*)$. The computer simulator encodes a function $\theta(x)$ that describes the relationship between some input parameters $x$ and a binomial distribution probability parameter $\theta(x)$. We suppose that there is a true, observable input value $X$, observable in the sense that, in theory data could be obtained to estimate $X$ directly, independently of the simulator. (To clarify, we have $x$ as an arbitrary choice of input value, and $X$ as the true, unknown values of the input quantities in reality.) Relating the simulator to reality, we recognise that the simulator is not perfect, so that $\theta^*=\theta(X)+\delta$, where $\delta$ represents the simulator error or discrepancy. The calibration problem is to infer $X$ given $Z$.

\subsection{Calibrating a stochastic computer simulator}

The computer simulator does not actually output $\theta(x)$ for a given input $x$. Instead, the simulator outputs a random variable $Y(x)$ with $Y(x)|\theta(x),n(x) \sim Bin(n(x),\theta(x))$. The value of $n(x)$ is expected to increase with the patient cohort size, the original patient sample size chosen for the simulator, but is subject to some random variation. Hence, for any simulator run at input $x$, we will have to infer the value of $\theta(x)$ based on the observations for $Y(x)$ and $n(x)$. During the calibration process, we will run the simulator at inputs $x_1,\ldots,x_m$, to obtain simulator data $D=\{x_i,Y(x_i),n(x_i)\}_{i=1}^m$, and so the aim of the calibration is to derive the posterior distribution $\pi(X|Z,D)$; we infer $X$ given $Z$ \textit{and} $D$.

We can evaluate the likelihood $\pi(Z\, |\, X=x, D)$ for $X$ at the value of $x$ via
$$
\pi(Z\, | \,X=x,D)=\iint \pi(Z|\theta(x),\delta,x,D)\pi(\theta(x)|x,D)\pi(\delta|\theta(x),x,D)d\theta(x) d\delta,
$$
which we assume can be simplified as
\begin{equation*}
\pi(Z \, | \, X=x,D)=\iint \pi(Z|\theta(x),\delta)\pi(\theta(x)|D)\pi(\delta|\theta(x))d\theta(x) d\delta.\label{outline-likelihood}
\end{equation*}
We make a further simplification: we suppose that we have run the simulator at $x$ to observe $Y(x)$ and $n(x)$, so that $\{x,Y(x),n(x)\}\in D$ and then we set 
$$
\pi(\theta(x)|D)=\pi(\theta(x)|Y(x),n(x)),
$$
so that we only use the run at $x$ to infer the corresponding $\theta(x)$. 

\subsection{Incorporating simulator discrepancy}


As we have already discussed, specifying a single choice of discrepancy distribution is difficult, and so we propose the following strategy to account for simulator discrepancy. We start with a conservative prior distribution for $\delta$ that permits moderately large values.  We obtain a sample from the posterior distribution $\pi(X|Z,D)$. We can then explore alternative distributions for $\delta$, using importance sampling to re-weight the sample according to alternative prior distributions $\pi(\delta)$. For example, in the case where $\delta$ is multivariate, corresponding to a multiple output simulator, we can investigate scenarios where some outputs are believed to be better modelled than others. By starting with a conservative prior for $\delta$ we are, in effect, `broadening the search' for inputs that give simulator outputs that are close to the observed data. Without any discrepancy, it is possible that no input value will give a good fit to all the output data.

Writing $\theta^*=\theta(X)+\delta$, we want the discrepancy term $\delta$ to add uncertainty about $\theta^*$ given $\theta(X)$, as we don't believe that running the simulator at the true observed $X$ (and an infinitely large cohort of patients) would give us $\theta^*$. To simplify the computation, we can achieve this effect by inflating the uncertainty about $\theta(x)$ given $n(x)$ and $Y(x)$, rather than by introducing an additional term $\delta$. We choose a $U[0,1]$ prior distribution for $\theta(x)$  and suppose that
\begin{equation*}
\theta(x)|Y(x),n(x) \sim Beta(1+\lambda Y(x),1+\lambda(n(x)-Y(x))),\label{theta-distribution}
\end{equation*}
with $\lambda \in (0,1]$. The parameter $\lambda$ has the effect of allowing for simulator discrepancy, by downweighting the information that the simulator run gives us about $\theta(x)$. In section \ref{sens-disc} we investigate the sensitivity of calibration to different choices of $\lambda$. Using the distribution for $\theta(x)$ given in section \ref{theta-distribution}, we assume $\delta=0$ and re-write the likelihood as
\begin{align}
\nonumber \pi(Z \, | \, X=x,D)&=\int  \pi(Z|\theta(x))\pi(\theta(x)|Y(x),n(x))d\theta(x)\\ 
&=\frac{^NC_ZB(1+\lambda Y(x)+Z, 1+\lambda(n(x)-Y(x))+N-Z)}{B(1+\lambda Y(x), 1+\lambda(n(x)-Y(x)))}, \label{outline-approx-likelihood}
\end{align}
where $B(.,.)$ is the Beta function.

\subsection{Sampling from the posterior distribution of the inputs} \label{sample-post}

Obtaining $Y(x)$ and $n(x)$ is computationally expensive, so we need to be selective in when we choose to run the simulator and evaluate the likelihood. We use importance sampling, where we construct a cheap-to-evaluate importance density using a Gaussian process emulator \citep{sacks}. In related works, \cite{ras} use a Gaussian process approximation to a (log) posterior density function to improve the efficiency of Bayesian integration,  which is extended in \cite{fielding} to include parallel tempering to accommodate multi-modality. Alternatively, \cite{bliz} use radial basis functions to provide a cheap-to-evaluate density function approximation. Constructing the emulator will be an iterative procedure, as the initial design region for the inputs may be specified somewhat conservatively, so that it may take several attempts to construct a satisfactory importance density. The outline procedure is as follows.

\begin{enumerate}
\item Using an initial set of simulator runs, investigate the design region to see if any subregions can be ruled out as having relatively low likelihood.
\item Run the simulator at a moderate number of input values over the reduced design region, to get initial simulator data $D=\{x_i,Y(x_i),n(x_i)\}_{i=1}^m$. Evaluate the likelihood in equation \eqref{outline-approx-likelihood} for each input value $x_i$.
\item Using the data from Step 2, construct a fast approximation of the log-likelihood using a Gaussian process emulator.
\item Construct an importance density for $\pi(X|Z)$ by approximating the log-likelihood by the posterior mean of the emulator. Use MCMC to generate a sample of values $X_1,\ldots,X_r$ from this approximate posterior density. To guard against the support of the importance density being too small, flatten the log-likelihood by multiplying it by a suitable constant.
\item Run the simulator at  $X_1,\ldots,X_r$ and evaluate likelihood (\ref{outline-approx-likelihood}) for each of these points. Calculate importance weights for each input.
\item If a small proportion of the inputs in $X_1,\ldots,X_r$ have relatively large importance weight, update the emulator to include the likelihood evaluations, and return to Step 4.
\end{enumerate}

\section{Calibration of a Natural History Model} \label{nhm}

\subsection{Natural History Models} \label{nhm-sim}

The basic set-up of the NHM is as follows; for a fuller description see \cite{pt-thesis}. The NHM represents a \emph{birth cohort}: a fixed-size sample of people followed from birth to death. A person in the cohort is deemed to have developed bowel cancer when they have reached the first cancer state, Duke's A, having begun in a non-cancer state, and progressed through three, ordered pre-cancer states: \mbox{low-,} medium- and high-risk adenomas. A person may continue to progress through three more increasingly severe cancer states, Duke's B, C and Stage D. Progression between states is governed by time. When in a given state, a progression time to the next state is simulated, together with a presentation time (the most common form of presentation being to visit a doctor), and a time until death. Out of these three actions, the one that occurs is the one with the shortest simulated time. Times are assumed to follow state-dependent Weibull distributions, the parameters of which form the majority of the NHM's unknown parameters that we calibrate. \texttt{}

This framework for a NHM allows a person's age to be known whenever they change state. It also allows a person to progress straight from birth to death (without ever contracting bowel cancer), or to progress through some or all pre-cancer and cancer states. By presenting a patient enters the health system where they receive a bowel cancer diagnosis. The age-based data that form part of the NHM's output result from these diagnoses and the tracking of ages. Having left the health system, a person returns to a non-cancer state and is still represented by the NHM, but their progression rates between states are elevated. While designed to mimic bowel cancer treatment within the health system, not all processes are necessarily well understood, or can be incorporated in the model. Simplifying assumptions, such as times following Weilbull distributions, are also required. These give examples of where discrepancy may arise.

The following gives details of the NHM's output required for calibration.

\subsection{Target data, output and notation} \label{app-calib}

The target data and NHM output are counts that we will in general denote by $Z_{jk}$ and $Y_{jk}(x)$, respectively, where $j = 1, \ldots, 4$ indexes the data type and $k = 1, \ldots, K_j$ indexes groups within types; corresponding sample sizes are denoted $N_{jk}$ and $n_{jk}(x)$, respectively. Here $x$ is the input vector that we use to initialise the NHM. The data types are identified explicitly, as opposed to considering the output as a single vector, due to their inherent differences, which will emerge in the following summaries.

\subsubsection{Cases by age} \label{agedata}

Target data $Z_{1k}$ represent a cross-sectional study and give the number of people out of $N_{1k}$ in the UK developing bowel cancer in 2008, where $k = 1, \ldots, 18$ indexes age groups 0-4, 5-9, $\ldots$, 80-84, 85+ \citep{crdata}. The NHM's output does not match the target data directly. Instead, it represents the cancer state and age of a birth cohort, ie. longitudinal data. To make the NHM output consistent with the target data, it is resampled by allocating each person to age group $k=1, \ldots, 18$ at random, according to probabilities determined by proportions in the UK population. Thus we take the NHM output, which corresponds to a longitudinal study, and resample it to match the target data, which corresponds to a cross-sectional study. Let $r=1, \ldots, R$ index each randomisation. The resulting NHM output corresponding to $Z_{1k}$ is denoted $Y_{1k}^{(r)}(x)$, with corresponding sample size $n_{1k}^{(r)}(x)$.  The likelihood
is obtained by averaging over randomisations, with $R$ large.

\subsubsection{Cases by type}

$Z_{2k}$ is the number of bowel cancer cases of type $k$ out of $N_2$ cases, where $k=1, \ldots, 4$ indexes types Duke's A, B and C, and Stage D, respectively. The NHM output is denoted $Y_{2k}(x)$ and is directly comparable to $Z_{2k}$. The total number of cases simulated is denoted $n_2(x)$.

\subsubsection{Obstructed cases by type}

These data also represent cases by type, but only those cases in which an obstruction (malignant large bowel) occurs and only for types Duke's B, C and Stage D \citep{obspaper}. We therefore define $Z_{3k}$, $N_3$, $Y_{3k}(x)$ and $n_3(x)$ similarly to $j=2$.

\subsubsection{Undetected adenomas by age}

$Z_{4k}$ is the number of people out of $N_{4k}$, where $k = 1, \ldots, 4$ indexes age groups under 55, 55-64, 64-74 and over 75, that had developed adenomas that had not been detected in their lifetime; these have later been detected in a necropsy study \citep{adepaper}. NHM output $Y_{4k}(x)$ and $n_{4k}(x)$ are defined similarly. 

\subsection{Discrepancy specification} \label{nhm-disc}

To introduce simulator discrepancy to the NHM, we consider reducing output sample sizes and counts, $n_{jk}(x)$ and $Y_{jk}(x)$, and specify these reductions as fractions, $\lambda_j \in (0, 1]$, $j=1, \ldots, 4$. We allow $\lambda$ to vary with data source because sample sizes in the NHM output vary in orders of magnitude. For example, the cases by age data are based only on those sample members that have developed cancer, whereas the undetected adenomas by age data are based on all patients in the model. To assess calibrated output, we consider its similarity to the target data, given approximate error bounds. These bounds represent how close a simulator output should be to the target data, considering three sources of error: sampling variability in the data, stochastic variability of the simulator output, and simulator discrepancy. For brevity, we present results for binomial data, though only minor alterations are required for multinomial data. 

We first consider error due to sampling variability. If $Z|\theta^* \sim Bin(N, \theta^*)$, then the variance of $p := Z/N$, which is used to estimate $\theta^*$, is $p(1-p)/N$. Similarly, if $Y(x) \sim Bin(n(x), \theta(x))$ is simulator output without discrepancy, the estimator $p(x) := Y(x)/n(x)$ has variance $p(x)(1-p(x))/n(x)$. The addition of simulator discrepancy, through $\lambda \in (0, 1]$, inflates the variance of the estimator to $p(x)(1-p(x))/(\lambda n(x))$, which can be partitioned as \[\frac{p(x)(1-p(x))}{\lambda n(x)} = \frac{p(x)(1-p(x))}{n(x)} + \frac{p(x)(1-p(x))(1-\lambda)}{\lambda n(x)}.\] Thus we decompose the variance of the simulator output into contributions due to the simulator being stochastic and that added by it being imperfect. We assess the calibrated output against the target data by considering approximate 95\% intervals around the target data, which widen as we add in the different sources of error: \begin{equation} \label{vardecompeq} \begin{array}{ll} \begin{array}{l}\text{measurement error} \\ \text{} \end{array} & \pm 2 \sqrt{\dfrac{p(1-p)}{N}},\\ \begin{array}{l}\text{measurement error and}\\ \text{simulator uncertainty} \\ \text{} \end{array} & \pm 2 \sqrt{\dfrac{p(1-p)}{N} + \dfrac{p(x)(1-p(x))}{n(x)}},\\ \begin{array}{l}\text{measurement error,}\\ \text{simulator uncertainty and} \\ \text{simulator discrepancy}\end{array} & \pm 2 \sqrt{\dfrac{p(1-p)}{N} + \dfrac{p(x)(1-p(x))}{n(x)} + \dfrac{p(x)(1-p(x))(1-\lambda)}{\lambda n(x)}}.\end{array} \end{equation} While $p(x)$, $n(x)$ and $Y(x)$ vary with $x$, they are estimated only once, from the simulator run with highest likelihood.

\begin{figure}[t]
\begin{center}
\includegraphics[width = 0.9\textwidth]{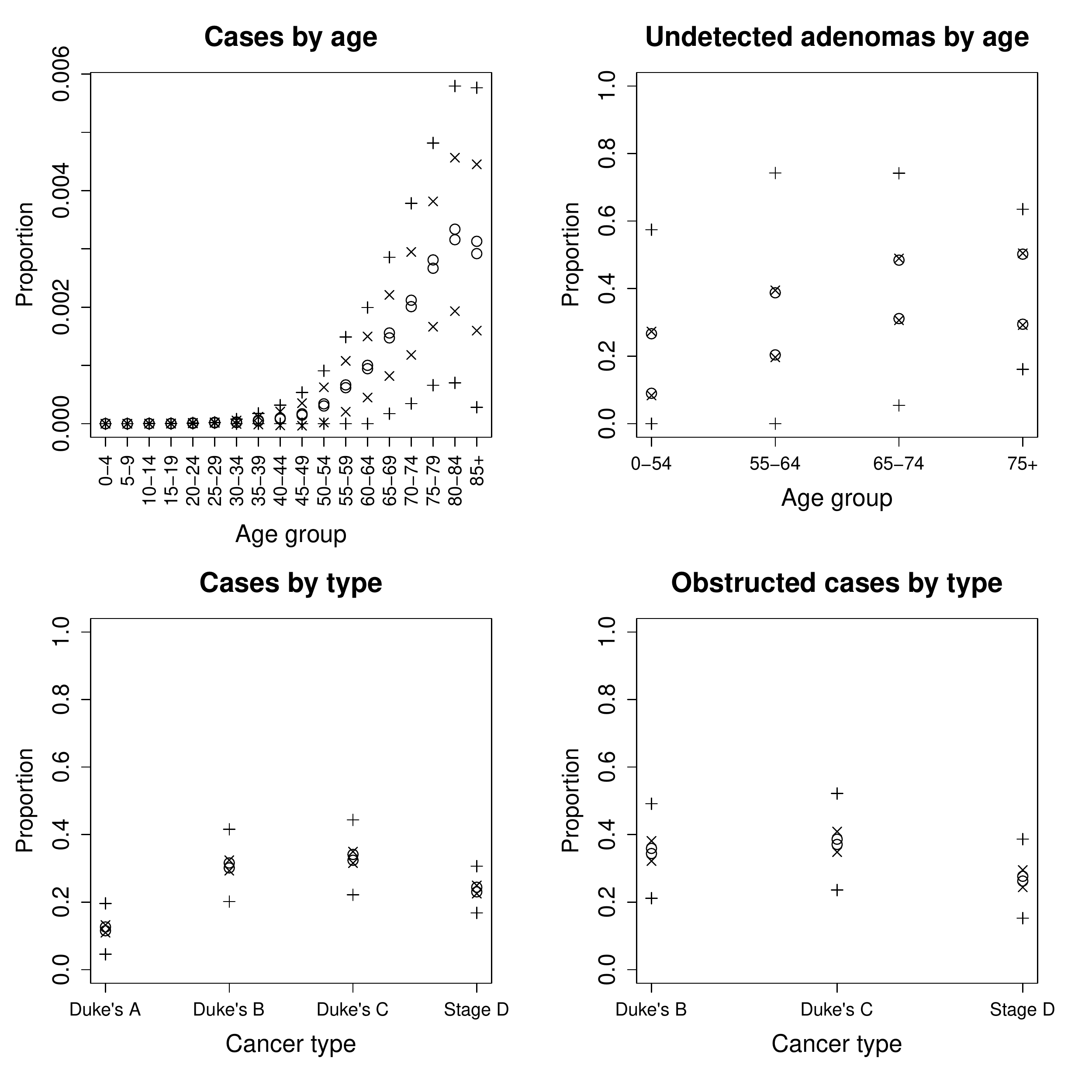}
\caption{\label{vardecomp}Variance decompositions for each target data source as described in section \ref{app-calib}. Cumulative contributions to variability (as given in equation set \eqref{vardecompeq}) due to target data ($\circ$), simulator uncertainty ($\times$) and simulator discrepancy ($+$) are shown.}
\end{center}
\end{figure}

Figure \ref{vardecomp} displays variance decompositions for each data source\footnote{Note that where proportions are all non-zero, representation on the logit scale might be more informative.}. This visual representation allows us to choose values of $\lambda_j$ `by eye': we choose values to give bounds around the target data that are such that, if output falls within the bounds, then we are prepared to deem it and its corresponding input plausible. We perform the calibration in waves and, in the build-up to the final calibration, can broaden the search for inputs by extending these intervals. We investigate sensitivity to different choices of $\lambda$ in section \ref{sens-disc}. In particular, our method is intended to make such sensitivity analyses relatively straightforward. Ultimately we set $\lambda_1 = 0.8$, $\lambda_2 = 0.008$, $\lambda_3 = 0.04$ and $\lambda_4 = 0.0004$, which are the values represented in Figure \ref{vardecomp}. Note that it is difficult to interpret the absolute value of the $\lambda_j$s, due to the different corresponding sample sizes generated internally in the model. We instead use Figure \ref{vardecomp} as the main tool for understanding how much discrepancy has been incorporated, and we later inspect the calibrated model outputs to assess how well the model can fit each type of data (see Figure \ref{summout}).

\subsection{Prior distributions for the calibration inputs}

The prior distributions for the inputs were independent uniform, set with conservatively wide ranges. It is possible that more carefully
specified priors would remove the need for some of the early waves in the history matching process (see section \ref{likcanc}). However, the elicitation problem would be hard, as the inputs do not all correspond to simple observable quantities. In that case, one might consider constructing a proper prior using the technique of `probabilistic inversion' \citep{du}, in which experts make judgements about model outputs, from which priors for model inputs are constructed. But the problem then would be that the experts may have already seen the calibration data, and may be unable/unwilling to provide judgements that do not take into account the known output data.

\subsection{Likelihoods for the cancer data} \label{likcanc}

Combining sections \ref{outline} and \ref{app-calib} allows us to calculate the likelihood for all the NHM's output. Notation for realisations follows from section \ref{app-calib}; for example, $z_{1k}$ is the observed number of people in age group $k$ developing bowel cancer out of $N_{1k}$ and $y_{1k}(x_i)$ is the corresponding NHM count out of $n_{1k}(x_i)$ for input $x_i$, with notation for other data types defined similarly. We model the cases by age and undetected adenomas by age data as binomially distributed, and assume weak prior information for its parameters by adopting a Uniform[0,1] prior. (Note that if population age-group proportions changed considerably over time, then the cases by age data could be subject to greater-than-binomial variation.) We assume that the cases by type and obstructed cases by type data are multinomially distributed, and use a Dirichlet({\bf 1}) prior to again represent weak prior knowledge. Finally the complete target data are $z = (z_1, z_2, z_3, z_4)$ where $z_j = (z_{j1}, \ldots, z_{jK_j})$. 


The overall log-likelihood for the complete target data for an input $x_i$ at which we have run the simulator and obtained output $y(x_i)$ is given by \begin{equation} \log\{\pi(z\,|\,X=x_i,\,y(x_i))\} = \sum_{j = 1}^4 \log(\pi_j), \label{likeq} \end{equation} where \begin{align*} \pi_1&= \dfrac{1}{R} \sum_{r = 1}^{R} \Bigg\{\prod_{k = 1}^{K_1} \dfrac{N_{1k}! \, B\big(1 + z_{1k} + \lambda_{1k} y_{1k}^{(r)}(x_i), \, 1 + N_{1k} - z_{1k} + \lambda_{1k}\{n_{1k}^{(r)}(x_i) - y_{1k}^{(r)}(x_i)\}\big)}{(N_{1k} - z_{1k})!\,z_{1k}!\,B\big(1 + \lambda_{1k} y_{1k}^{(r)}(x_i), \, 1 + \lambda_{1k}\{n_{1k}^{(r)}(x_i) - y_{1k}^{(r)}(x_i)\}\big)}\Bigg\} \intertext{with index $r=1, \ldots, R$ denoting the $r$th randomisation of the NHM output,} \pi_2 &= \dfrac{N_2!\, \{\lambda_2 n_2(x_i) + K_2 - 1\}!}{\{N_2 + \lambda_2 n_2(x_i) + K_2 - 1\}!}\prod_{k=1}^{K_2}\dfrac{\{z_{2k} + \lambda_2 y_{2k}(x_i)\}!}{z_{2k}!\{\lambda_2 y_{2k}(x_i)\}!},\\ \pi_3 &=\dfrac{N_3!\, \{\lambda_3 n_3(x_i) + K_3 - 1\}!}{\{N_3 + \lambda_3 n_3(x_i) + K_3 - 1\}!}\prod_{k=1}^{K_3}\dfrac{\{z_{3k} + \lambda_3 y_{3k}(x_i)\}!}{z_{3k}!\{\lambda_3 y_{3k}(x_i)\}!} ,\\ \pi_4&= \Bigg\{\prod_{k = 1}^{K_4} \dfrac{N_{4k}! \, B\big(1 + z_{4k} + \lambda_{4k} y_{4k}(x_i), \, 1 + N_{4k} - z_{4k} + \lambda_{4k}\{n_{4k}(x_i)  - y_{4k}(x_i)\}\big)}{(N_{4k} - z_{4k})!\,z_{4k}!\,B\big(1 + \lambda_{4k} y_{4k}(x_i), \, 1 + \lambda_{4k}\{n_{4k}(x_i) - y_{4k}(x_i)\}\big)}\Bigg\}.\end{align*}

\begin{figure}[t]
\begin{center}
\includegraphics[width = \textwidth]{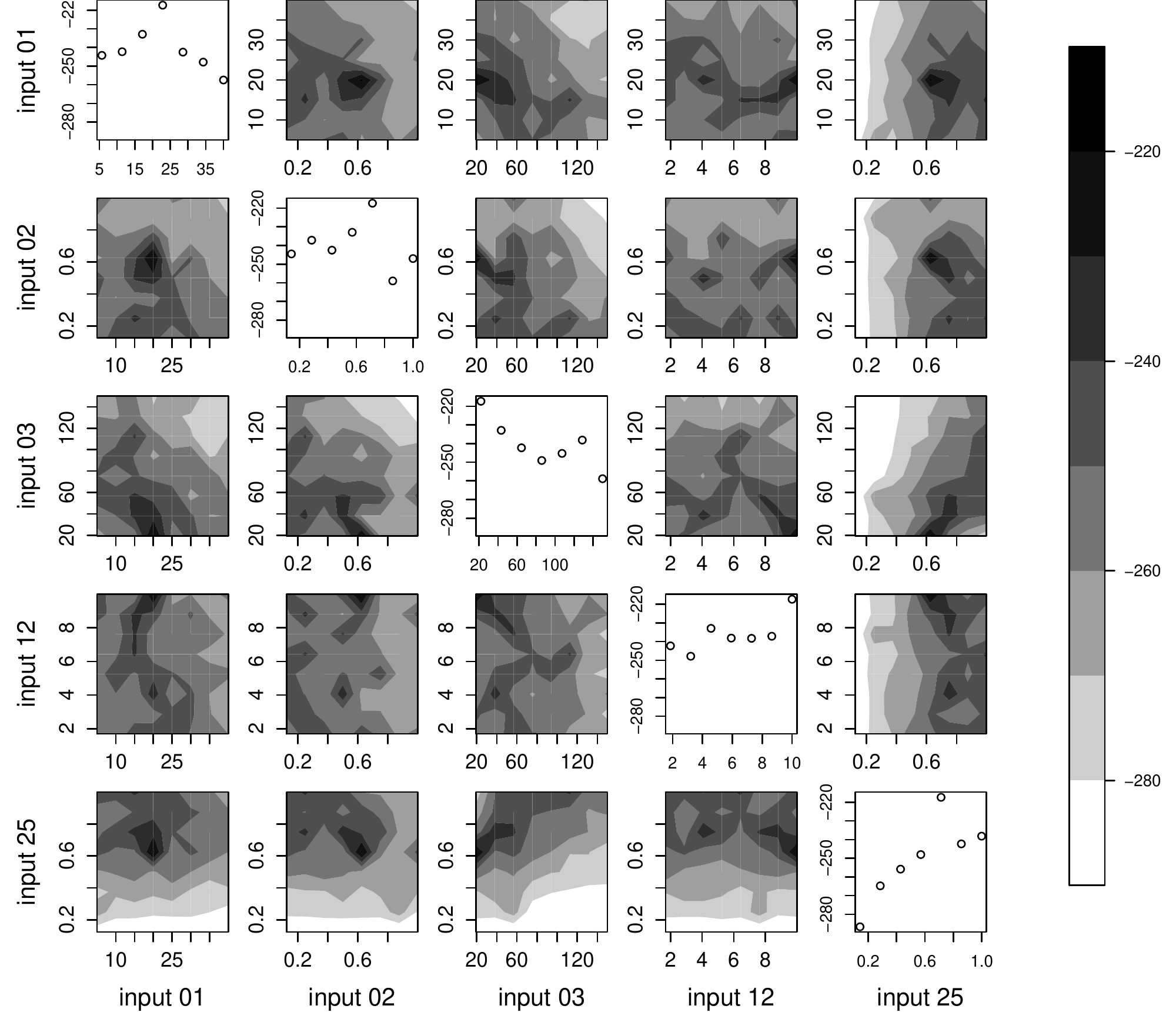}
\caption{\label{inputreg}Pairwise maximised log-likelihood (off-diagonal) and marginal binned maximised log-likelihoods (diagonal) for inputs 1, 2, 3, 12 and 25. (Pairwise plots are a smoothed representation of an $8 \times 8$ grid.)}
\end{center}
\end{figure}

We calculate the log-likelihood for 10,000 NHM runs, each using a birth cohort of size 100,000. Figure \ref{inputreg} shows the log-likelihood against inputs 1, 2, 3, 12, and 25, specifically against single inputs (achieved by maximising the likelihood over equal-sized bins) and for pairwise combinations of inputs (achieved by maximising over grid cells). Input 1 represents the age at which a person can develop adenomas, input 2 the log-parameterised Weibull shape parameter for progression times between pre-cancer states, input 3 the Weibull scale parameter for progression to the first pre-cancer state, input 12 the change in Weibull scale parameters due to having previously been treated for cancer and input 25 the probability that a person develops adenomas in their lifetime.

Figure \ref{inputreg} shows that for some regions of input space the log-likelihood is much higher than for others. We use where the likelihood is relatively high to define a reduced input space, which is specified by marginal ranges and pairwise regions. Because we start with broad parameter ranges for all 25 inputs, there is large variation in the likelihood values of Figure \ref{inputreg}, and so our criterion
for ruling out parts of input space is set conservatively: we omit parts where the likelihood ratio, relative to the observed maximum, fails to exceed e$^{-40}$. This reduces the input space to 0.7\% of its original size. As we approximate true maximum log-likelihoods by those observed, we make conservative choices here to compensate for observed maxima being underestimates of the true maxima. This could be avoided if it were feasible to use many more simulator runs. The technique of reducing the input region is related to that used in history matching by \cite{vern-galaxy}, in which implausibility of parts of input space is quantified, and parts measured to have large implausibility are ruled out.

As in \cite{vern-galaxy}, the input region can be further reduced in waves. Here second and third waves, also of 10,000 NHM runs, are performed, which use birth cohorts of 200,000 and 300,000 people, respectively. It is possible that, when reducing the input region, more carefully specified priors could remove the need for some of these early waves. The emulator training data are based on the region chosen after the third wave, which is 0.0001\% the size of our starting input region.  

\subsection{Emulator specification and building} \label{em-build}

We are building an emulator for the function $f(x)$, the log-likelihood for input vector $x$, where $x=(x^{(1)}, \ldots, x^{(p)})^T$, which is defined in section \ref{likcanc} as \[f(x) := \log\{\pi(z\,|\,X=x,\,y(x))\} = \sum_{j=1}^4 \log \pi_j.\] Thus we model \[f(x)\, | \, \sigma^2, \beta, \phi, \nu^2 \sim GP(h^T(x) \beta, \sigma^2 c(x, \,))\] where $h(\,)$ and $\beta$ comprise $q$ basis functions  and regression coefficients, respectively, $h^T(x) \beta$ is therefore the GP mean function, $\sigma^2$ is its variance and $c(\, , \,)$ is its correlation function. 

We choose the correlation function to have the Gaussian form \begin{equation*} c(x_i, x_j) = \left\{\begin{array}{ll} \exp\{ -\sum_{d = 1}^p ((x_i^{(d)} - x_j^{(d)}) / \phi_d)^2\} & \text{if}~~ x_i \neq x_j,\\(1 + \nu^2/\sigma^2)^{-1} & \text{if}~~ x_i = x_j,\end{array} \right. \label{corstruc} \end{equation*} for a set of roughness parameters $\phi = \{\phi_1, \ldots, \phi_p\}$, where $\phi_d > 0$, $d=1, \ldots, p$. The parameter $\nu^2 > 0$ introduces a \emph{nugget} effect into the emulator, which has been shown to improve the predictive performance of Gaussian process emulators \citep{andri-chal, gram-lee}, but is imperative for a stochastic simulator such as the NHM. We are prepared to accept a constant nugget on the grounds that ultimately it is the emulator's posterior mean that we use to sample inputs. The nugget effect could be allowed to vary with the inputs, but any functional form for this relationship is not obvious, and while we investigated some log-linear forms, none improved upon the constant choice. We choose the Gaussian form because we expect the underlying function to be smooth, and the inclusion of the nugget term is likely to make the precise choice less critical, as we are not trying to interpolate the training data exactly. 

The emulator is specified to have a constant mean function, ie. $h(x) =1$. This choice is convenient here because many runs have very low likelihood, which results in a small mean for the Gaussian process, and consequently no inputs being sampled far away from those with a high corresponding likelihood. Polynomial terms could be added. We tested a linear form, but this gave unsatisfactory results, as inputs far away from those with simulator runs would be sampled if they had a high value of the linear predictor. A quadratic form with interactions might combat this, but as the NHM has 25 inputs, this was impractical. Perhaps more suitable would be (the log of) a parametric density function, though this gives a mean function that is non-linear in its parameters. 


We use 2,000 simulator runs for the emulator training data, which are chosen using a Maximin Latin hypercube design on the reduced region chosen after wave three in section \ref{likcanc}. We define the following: input set $D_X=\{x_1, \ldots, x_m\}$; vector of corresponding log-likelihoods $f(D_X) = \big(f(x_1), \ldots, f(x_m)\big)^T$; $m \times m$ matrix $A$, which has $(i, j)$th element $c(x_i, x_j)$; and $t(x)^T = (c(x_1, x), \ldots, c(x_m, x))$. 

For the hyperparameter prior we choose $\pi(\sigma^2,\, \beta,\, \phi,\, \nu^2) \propto \sigma^{-2}$. It follows that posterior emulator is given by \[f(x)\, | \, D, \phi, \nu^2 \sim tP_{n-q}(\hat \beta, \hat \sigma^2 c^*(x, \,)),\] a Student $t$-process on $n-q$ degrees of freedom, where \begin{align*} \hat \beta &= (1_m^T A^{-1} 1_m)^{-1} 1_m^TA^{-1} f(D_X)\\  \hat \sigma^2 &= (m - q - 2)^{-1} (f(D_X) - \hat \beta)^T A^{-1} (f(D_X) - \hat \beta) \\ m^*(x) &= \hat \beta + t(x)^T A^{-1}\big(f(D_X) - \hat \beta\big) \shortintertext{and} c^*(x,\,x') &= c(x,\,x') - t(x)^TA^{-1}t(x') \label{vstar} \\& \hspace{1cm} +\big(1 - t(x)^TA^{-1}1_m\big)(1_m^T A^{-1} 1_m)^{-1}\big(1 -  t(x')^TA^{-1}1_m\big)^T. \nonumber
\end{align*} Finally, $(\phi,\,\nu^2)$ has posterior \begin{equation*} \pi^*(\phi,\,\nu^2) \propto (\hat \sigma^2)^{-(m - q)/2} |A|^{-1/2} |1_m^TA^{-1}1_m|^{-1/2} \, \pi(\phi,\,\nu^2). \label{hatphi}\end{equation*}  We fix $(\phi, \nu^2)$ at the mode of $\pi^*(\phi,\,\nu^2)$. This is found using the Nelder-Mead optimisation algorithm, which is initialised with 200 iterations of the Gibbs sampler, in which Metropolis-Hastings updates are used.

\subsection{Input sampling} \label{input-sampling}

For the algorithm of section \ref{sample-post} to perform well, the emulator should represent high values of the log-likelihood fairly accurately. We use importance sampling to give a sample of inputs, and for the importance density use the emulator posterior mean, which serves as an approximation to the log-likelihood. We can sample from the importance density by again using Gibbs sampling with Metropolis-Hastings updates. To obtain the calibrated inputs we identify parts of the input region where the difference between the posterior mean and the log-likelihood is large, or where, given the posterior mean is relatively large, the emulator's uncertainty is large. The latter is identified using the pivoted Cholesky decomposition \citep{pivchol}. We can then add simulator runs in these parts to enable the emulator to provide a more accurate representation of the log-likelihood surface. The following algorithm then describes how we obtain the final sample of calibrated inputs. 

\begin{enumerate}
\item Obtain a sample of inputs, $D_S = (X_1, \ldots, X_S)$, by Gibbs sampling using the emulator posterior mean, $m^*(x)$, to approximate the log-likelihood. \label{start}
\item Compute the pivoted Cholesky decomposition of the covariance matrix for the sample, ie. the $S \times S$ matrix $A_S$ with $(i,j)$th element $c(X_i, X_i)$, $i, j = 1, \ldots, S$, and let $\{p_s\}_{s=1}^S$ denote its diagonal elements. Sort $D_S$ by the pivot, and take the first $u$ members, to give $D_{piv}$, where $u$ is the maximum number of simulator runs we are prepared to add to the training data in one iteration.
\item Define $p_s$ to be `large' if $p_s>v$, for some $v>0$. If no $p_s$ are large, proceed to Step \ref{start3}. Otherwise form the set $D^\dag = \{X_s \in D_{piv}\,:\,p_s>v\}$, for $s=1, \ldots, u$, evaluate the simulator at each of its members and calculate their log-likelihoods, $f(D^\dag)$. \label{st1}
\item Add $D^\dag$ and $f(D^\dag)$ to the training data, re-build the emulator, and return to Step \ref{start}. \label{start2}
\item Compute importance weights $w_s = \exp\{f(X_s) - m^*(X_s)\}$ for $X_s \in D_S$. If a large proportion of weights are zero, return to Step \ref{start2}. \label{start3}
\item Obtain the calibrated inputs, $D^* = \{X_1^*, \ldots, X_M^*\}$, by resampling $D_S$ with replacement according to weights $w_s^*= w_s/ \sum_{s=1}^S w_s$. 
\end{enumerate}

\begin{figure}[h!]
\begin{center}
\includegraphics[width=0.9\textwidth]{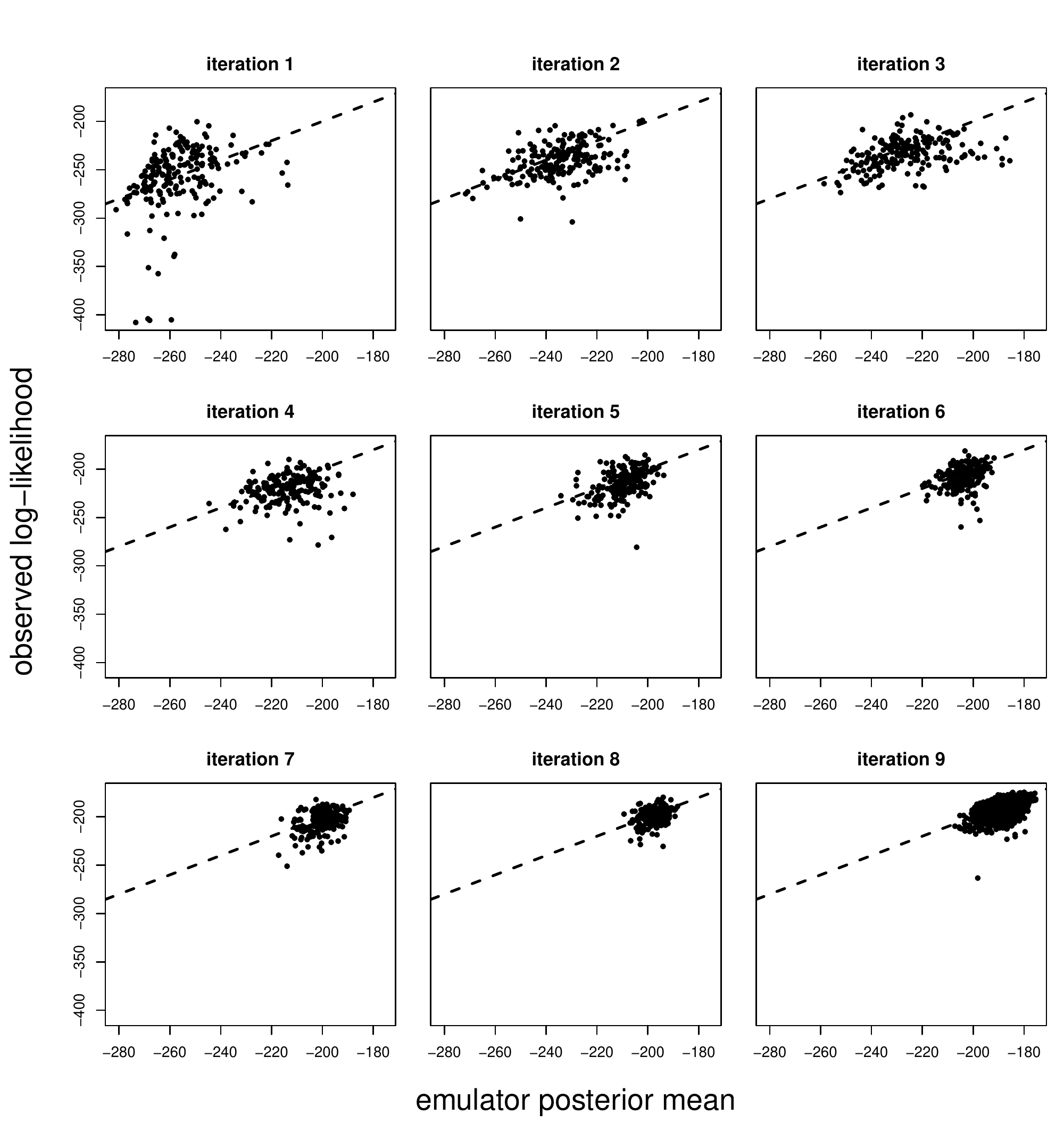}
\caption{\label{likvsem}Observed log-likelihoods against emulator posterior means (based on previous iteration) at iterations 1--8 for samples of size 200 and iterations 9 for a sample of size 1000. The line $y=x$ is superimposed ( - - - ).}
\end{center}
\end{figure} 

For Step \ref{start} of the calibration algorithm we choose $S=2,000$, which is achieved by thinning an initial sample of size 100,000 by 50. For Step 2 we choose $u=200$ and for Step \ref{st1} choose $v=2$. During the first iteration of the algorithm we find that almost all $p_s$ are large, which indicates that the emulator's uncertainty is large for all the sampled inputs. Consequently, the importance density may have insufficient support where the true log-likelihood is high. We flatten the log-likelihood to compensate for this, which is achieved by using $\alpha m^*(x)$ instead of $m^*(x)$, $0 < \alpha \leq 1$, in Step \ref{start}; we initially choose $\alpha=0.1$. Introducing $\alpha$ can also combat multi-modality of the log-likelihood, as found for parallel tempering in \cite{fielding}. Log-likelihoods calculated for the simulator runs are then compared against previous emulator posterior means, that is comparing $f(x)$ with $E(f(x)\,|\,D)$ for $x \in D^\dag$, where $D$ are the last-used training data. This comparison is shown for iterations 1--9 in Figure \ref{likvsem}. 

\begin{figure}[h!]
\begin{center}
\includegraphics[width=0.9\textwidth]{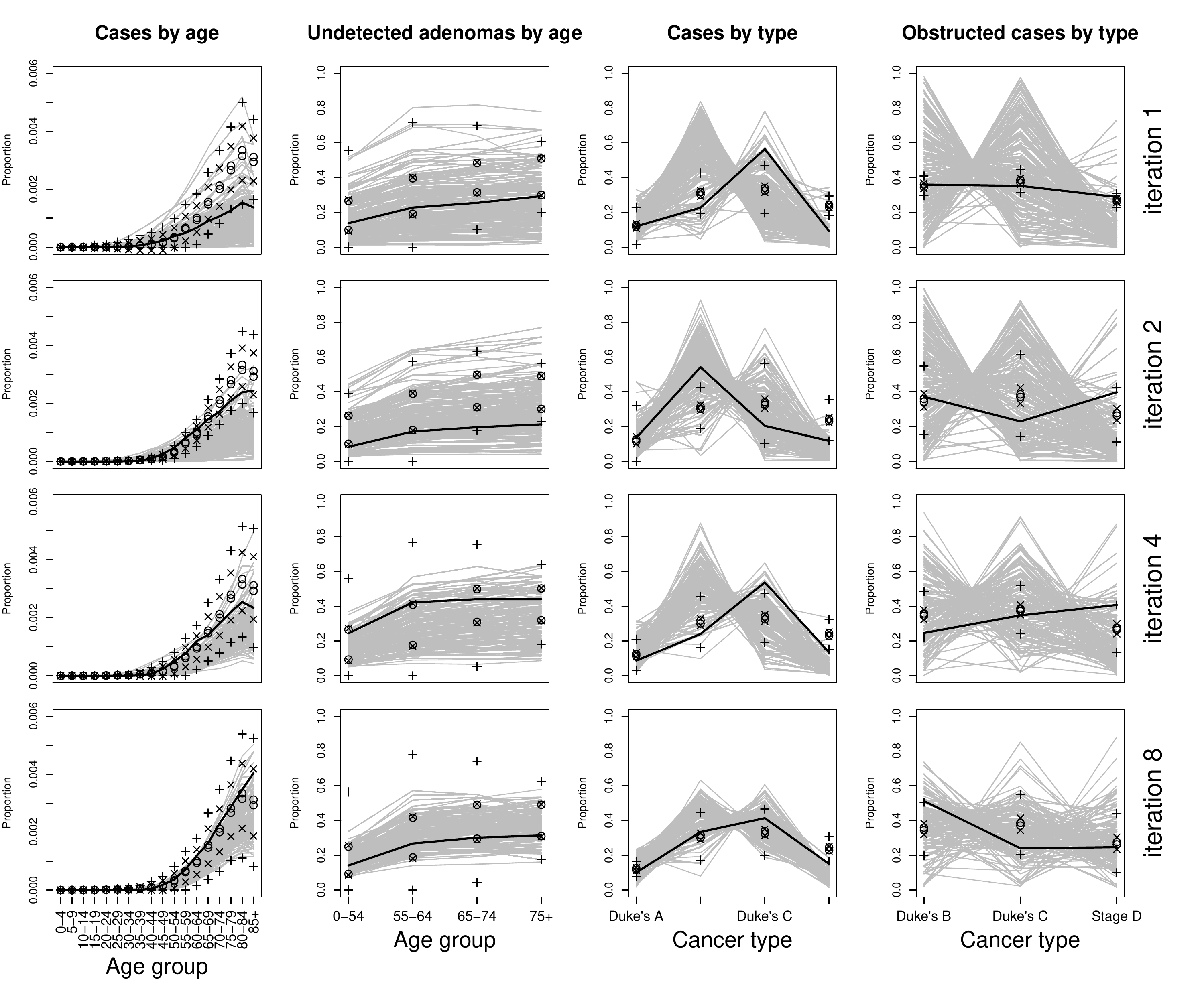}
\caption{\label{itsum}NHM output against target data for iterations 1, 2, 4 and 8. Uncertainty bounds are as in Figure \ref{vardecomp}. The black line highlights the run with highest likelihood.}
\end{center}
\end{figure}

From Figure \ref{likvsem}, we see that the agreement between $f(x)$ and $E(f(x)\,|\,D)$ is poor for the first iteration, which means that the emulator posterior mean will not serve well as an importance density for sampling inputs from the log-likelihood. We also look at how the simulator's output compares with the target data, given expected levels of uncertainty (as described in section \ref{nhm-disc}), which is shown for iterations 1, 2, 4, and 8 in Figure \ref{itsum}. For iteration 1, while some runs give a good match to some of the target data, most fail to provide an adequate match to all of the target data.

We proceed to perform further iterations. For iteration 2 we increase $\alpha$ to 0.2, and find that the match between $f(x)$ and $E(f(x)\,|\,D)$ has improved, but is still unsatisfactory, which can be seen in Figure \ref{itsum}. Therefore we perform further iterations, increasing $\alpha$ by 0.1 for each. Adequate agreement between the emulator and observed log-likelihoods is achieved by iteration 8, which is confirmed by iteration 9, the latter of which we choose to be the final emulator. There is some suggestion from Figure \ref{itsum} of disagreement between the NHM output and the target data at iteration 8; however, the points used to assess this are those for which the emulator's conditional variance is greatest, and therefore a better match between the emulator's posterior mean and the true log-likelihoods can be expected for a random sample of inputs. Furthermore, we only need approximate agreement between the emulator posterior mean and the true log-likelihood, because those points for which agreement is poor will be downweighted during importance sampling. Further iterations could instead be performed to improve agreement, but here that was found to be less efficient than having some negligible importance weights. We therefore deem the emulator to be adequate for providing a proposal distribution for the importance sampler.

\subsection{Calibrated output} \label{calib-out}

\begin{figure}[h!]
\begin{center}
\includegraphics[width=0.95\textwidth]{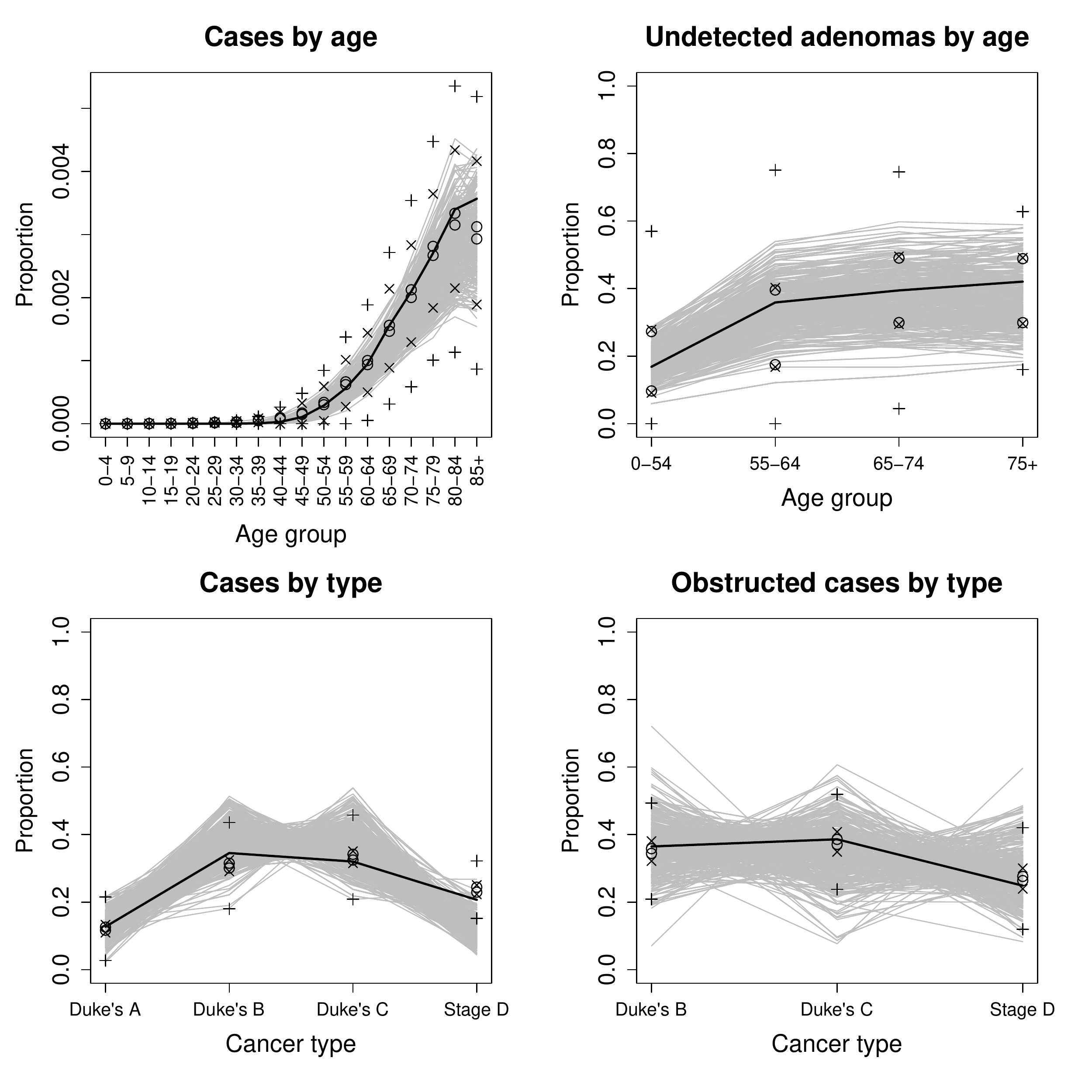}
\caption{\label{summout}Calibrated NHM runs against target data.}
\end{center}
\end{figure}

We use the emulator estimated at iteration 9 for the final sample of calibrated inputs. We choose this sample to be of size 1,000, and obtain it from an importance sample of size 2,000 by sampling with replacement according to the importance weights, ie. $\exp\{f(x) - m^*(x)\}$. Figure \ref{summout} shows the calibrated NHM output against the target data for the four different data types. We can see the calibration to have worked well, as the calibrated output is consistent with the target data, once we account for uncertainty amounts. 

\subsection{Sensitivity to the discrepancy specification} \label{sens-disc}

\begin{figure}[h!]
\begin{center}
\includegraphics[width=0.9\textwidth]{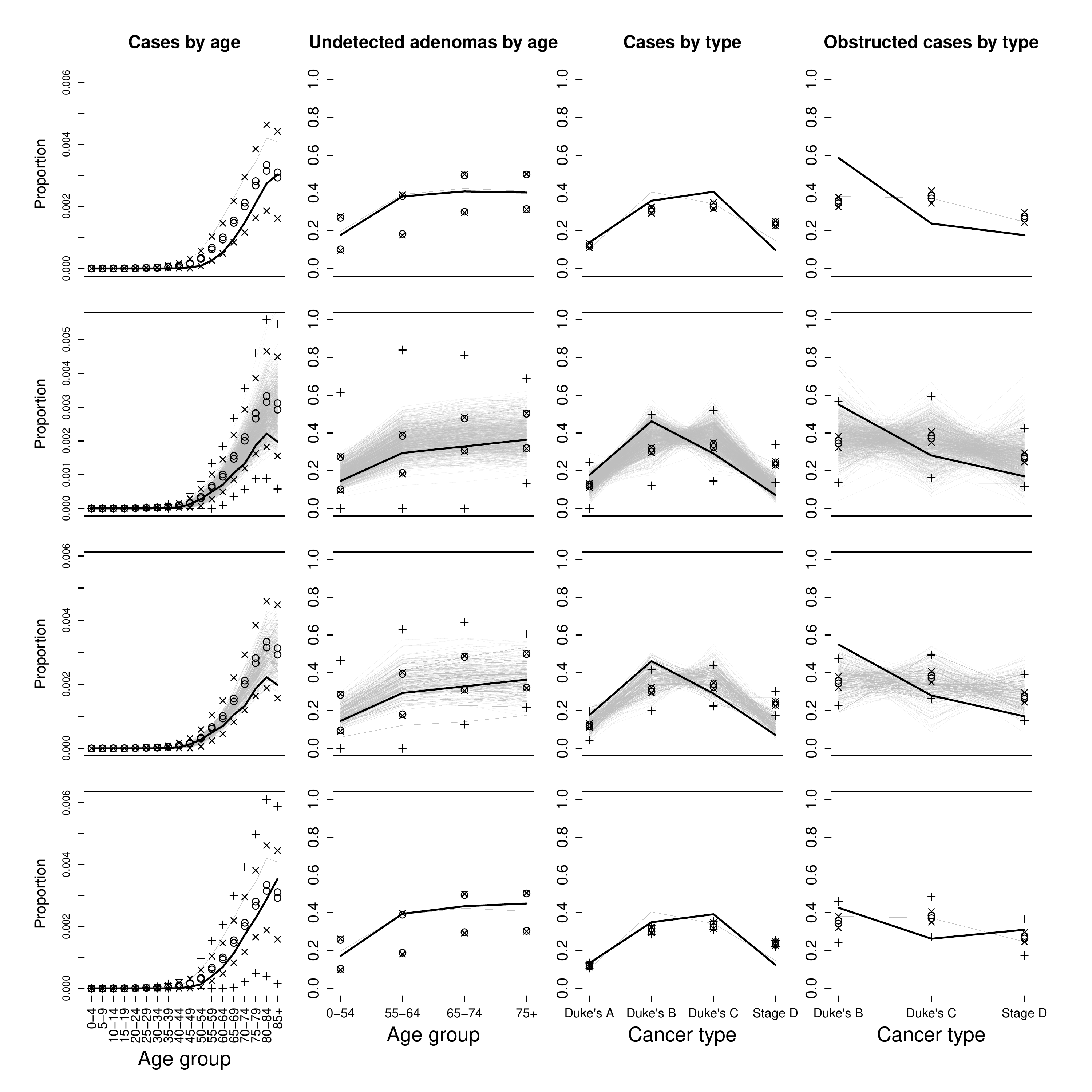}
\caption{\label{sumdisc}Summaries of simulator output against target data for various discrepancy specifications: no discrepancy for any data source (row 1), discrepancy levels doubled (row 2), no discrepancy for cases by age (row 3) and no discrepancy for cases by type (row 4).}
\end{center}
\end{figure}

We have calibrated the NHM using discrepancy values of $\lambda_1 = 0.8$, $\lambda_2 = 0.008$, $\lambda_3 = 0.04$ and $\lambda_4 = 0.0004$. We can investigate sensitivity to these choices by simply recalculating log-likelihoods and then importance weights for alternative discrepancy values. This requires little computational cost compared to re-running the simulator. The calibrated output for four alternative discrepancy specifications is shown in Figure \ref{sumdisc}. 

In the first of these alternative discrepancy scenarios, we consider the case where no discrepancy is assumed, which would imply that the simulator is a perfect representation of reality at the true value of $X$. This results in an unsatisfactory calibration: all but two of the simulator runs have negligible importance weights, one of which is much larger than the other, and the output from neither of these runs matches the target data, given uncertainty amounts. We then consider doubling discrepancy amounts, relative to our preferred amounts, so that $\lambda_1 = 0.4$, $\lambda_2 = 0.004$, $\lambda_3 = 0.02$ and $\lambda_4 = 0.0002$. This results in the importance sample having a greater range, when compared to the original calibrated inputs of section \ref{calib-out}, and in turn gives more variability in the calibrated output. While altering the discrepancy specification has changed the distribution of the calibrated inputs, the change in distribution of corresponding output is relatively small, which suggests that we do not need to be overly precise when specifying the discrepancy in order to achieve a reliable calibration. 

We also consider assuming no discrepancy for only one data source, leaving discrepancy values for the remaining sources unchanged. If we assume no discrepancy for the cases by age data, then the calibrated output still matches the target data for the cases by age data and for the other data sources, and the sample of calibrated inputs also contains sufficiently many unique values. However, when we assume no discrepancy for the cases by type data, the sample of calibrated inputs returns to containing only two unique members (the same two as when no discrepancy is assumed for all data sources), and for cases by type the calibrated output fails to match the target data. In summary, though, we find that while discrepancy amounts need some consideration, the precision that specifications require is within our capabilities, allowing the NHM to be calibrated reliably. However, the calibration becomes unsatisfactory when we ignore discrepancy, or specify it poorly.

\section{Discussion} \label{discuss}

In this paper we have calibrated a Natural History Model so that its output is consistent with reality. However, in order to do this we have had to address three important issues that arise when calibrating the computer simulator. The first is calibrating a simulator of `moderate' computational expense, that is one for which calibration it is not practical using Monte Carlo simulation alone, but nor is it one that requires us to rely solely on a computationally cheap surrogate model, such as a Gaussian process emulator. We therefore propose a calibration method that may be thought of as a hybrid of the two, which uses an emulator to provide a preliminary, approximate calibration, and combines this with simulator run data, through importance sampling, to give a final and more accurate calibration. Because the simulator is only of intermediate computational expense, we have taken a conservative approach to calibration, which can be seen in the criteria for refining the design region (section \ref{likcanc}) and when we `flatten' the log-likelihood (section \ref{em-build}). Were the simulator more expensive, we might need to consider optimising the calibration process to minimise the number of simulator runs needed.

The use of importance sampling has allowed us to explore a further issue, which is the sensitivity of calibration to different discrepancy specifications, which is important to understand because discrepancy must be adequately quantified before we can calibrate a simulator \citep{bryn}. In particular, while we can in theory always adjust a discrepancy specification and check the sensitivity of a calibration to adjustment, in practice this is likely to be impractical due to its computational requirements. Here, though, such investigation becomes computationally feasible, as we simply need to recalculate importance weights and obtain a new sample of calibrated inputs in order to assess different discrepancy specifications. This does need the original importance sample to be suitable, in particular for it to have enough non-negligible importance weights under the new discrepancy specification.

Finally we address how to calibrate a simulator, which we already know to be of intermediate computational expense, that is stochastic and has output that contains count data. We achieve this by using a Gaussian process prior for the log-likelihood, as the log-likelihood is better suited to the Gaussian process assumptions than the simulator output itself. It also reduces the task of calibrating 30-dimensional output to one in which we only have to model a one-dimensional entity. Introducing a nugget effect, overcomes the simulator being stochastic, which will reflect in the log-likelihood surface. 

The motivation for the calibration is to support decision-making, and so the main objective for incorporating simulator discrepancy is to protect against over-confidence. Although we have incorporated discrepancy into the four output types, the analysis is less informative for understanding the causes of simulator error, and where simulator improvements would be beneficial. Our approach to discrepancy is also less suited to capturing systematic errors, which could arise from posterior correlation in the cases by age data (Figure \ref{summout}), but is not recognised in likelihood \eqref{likeq}. Such issues may be better addressed with the `internal' simulator discrepancy approach in \cite{strong}. Nevertheless, the present calibrated simulator, with allowance made for discrepancy, will still have significant value in supporting decisions.

\section*{Acknowledgements}
We thank Paul Tappenden for providing the NHM and for guidance on its usage, and thank two reviewers and an Associate Editor for suggestions that have brought improvement to this paper. This work was supported by RCUK funding for the \emph{MUCM2} project (grant EP/H007377/1).

\newcommand{\noopsort}[1]{}

\end{document}